\documentclass[12pt]{article}
\usepackage[mathscr]{eucal}
\usepackage{amsfonts}
\usepackage{amssymb}
\usepackage{amsthm}
\def\theequation{\arabic{section}.\arabic{equation}}
\newcommand{\be}{\begin{eqnarray}}
\newcommand{\ee}{\end{eqnarray}}
\newcommand{\ba}{\begin{array}}
\newcommand{\ea}{\end{array}}
\newcommand{\p}[1]{(\ref{#1})}
\newcommand{\lb}[1]{\label{#1}}

\def\bbox{{\,\lower0.9pt\vbox{\hrule \hbox{\vrule height 0.2 cm
\hskip 0.2 cm \vrule height 0.2 cm}\hrule}\,}}
\newcommand{\dsl}{\pa \kern-0.5em /}

\newcommand{\nn}{\nonumber \\}

\begin{document}


\begin{titlepage}

\vfill
\vfill

\begin{center}
\baselineskip=16pt {\Large  Real and Complex Supersymmetric $d=1$
Sigma Models
\vspace{0.2cm}

With Torsions} \vskip 0.3cm {\large {\sl }}
\vskip 10.mm {\bf ~S.~A. Fedoruk$^{*, 1, \star}$, ~E.~A. Ivanov$^{*, 2}$,
A.~V. Smilga$^{\dagger, 3}$}
 \\
\vskip 1cm
$^*$
Bogoliubov Laboratory of Theoretical Physics, \\
JINR, 141980 Dubna, Russia\\

\vspace{6pt}
$^\dagger$
SUBATECH, Universit\'e de Nantes, \\
4 rue Alfred Kastler, BP 20722, Nantes 44307, France;\\
On leave of absence from ITEP, Moscow, Russia

\end{center}
\vskip 2cm

\par
\begin{center}
{\bf ABSTRACT}
\end{center}
\begin{quote}

We derive and discuss, at both the classical and the quantum levels,
generalized ${\cal N} = 2$ supersymmetric quantum mechanical sigma
models describing the motion over an arbitrary real or an arbitrary
complex manifold with extra torsions. We analyze the relevant vacuum
states to make explicit the fact that their number is not affected by
adding the torsion terms.

\vfill
 \hrule width 5.cm
\vskip 2.mm
\noindent $^1$ fedoruk@theor.jinr.ru\\
\noindent $^2$ eivanov@theor.jinr.ru\\
\noindent $^3$ smilga@subatech.in2p3.fr\\
\noindent $^\star$ {\small On leave of absence
from V.N.\,Karazin Kharkov National University, Ukraine}\\

\end{quote}
\end{titlepage}
\setcounter{equation}{0}
\section{Introduction}

Sigma model is a theory where the configuration space on which the
dynamic variables are defined is not the flat space $\mathbb{R}^D$, but represents a nontrivial target
manifold of dimension $D$.
The number of physical space-time coordinates can vary. In the simplest case of mechanical system, all
variables depend only on time. In the absence of external fields, the bosonic
Lagrangian of such $d=1$ sigma model is then given by
 \be
\label{Lbos}
 L^{\rm bos} \ =\ \frac 12\, g_{MN}(x)\,  \dot{x}^M \dot{x}^N\,.
 \ee
It describes the free motion over the manifold with coordinates $x^M$, $M=1,\ldots,D\,$, and the metric $g_{MN}(x)$.

The Lagrangian (\ref{Lbos}) can be supersymmetrized in different ways, yielding, after quantization,
various versions of supersymmetric quantum mechanics (SQM).
One can, e.g., introduce D  real ${\cal N} = 1$  superfields
\footnote{${\cal N}$ counts the number of {\it real} supersymmetries.},
${\cal X}^M \ =\ x^M + i\theta \psi^M\, $, and write the action \cite{A-GFW}
    \be
\label{SN=1}
S &=&  \frac i2 \int d\theta dt \, g_{MN}({\cal X}) \, {\cal D}
{\cal X}^M\, \dot{{\cal X}}^N \nn
&=& \frac 12 \int \, dt  g_{MN}(x) (\dot{x}^M \dot{x}^N + i \psi^M \nabla \psi^N )\,,
    \ee
with $ {\cal D} = \partial_\theta - i\theta \partial_t$, $g_{MN}=g_{NM}$ and
$\nabla \psi^N = \dot{\psi}^N + \Gamma^N_{PQ}\dot{x}^P \psi^Q $.
 The quantum version of the corresponding supercharge can be associated
with the Dirac operator $/\!\!\!\!{\nabla}$.\
 \footnote{There exists also another quantum supercharge
 $\tilde Q  =  /\!\!\!\!{\nabla} \gamma^{D+1}$ where $\gamma^{D+1}
= \prod_A \gamma^A \equiv \prod_A \psi^A $ ($\gamma^A \equiv \psi^A
= \psi^M e_M^A$), i.e. the quantum Hamiltonian enjoys here the
${\cal N} = 2$ supersymmetry required to make the spectrum  double
degenerate. Note that the quantum supersymmetry algebra
\be
Q^2 =
\tilde{Q}^2 = H\, , \ \ \ \ \ \ \ \ \ \ \ \ \ \{Q, \tilde{Q} \}=
0
\ee
cannot be preserved at the classical level, because the
Poisson bracket $\{\tilde{Q}, \tilde{Q} \}_{P.B.}$ vanishes
\cite{Holten}. Thus, we are facing here an interesting phenomenon of
the {\it classical} anomaly of supersymmetry
 \cite{clasanom}
({\it quantum} anomalies when one cannot keep a classical Lagrangian
symmetry at the quantum level are, of course, much better known). }

Another possibility is to introduce the real ${\cal N} = 2$ superfields with twice as many fermion degrees of freedom,
\footnote{This ${\cal N}=2, d=1$ multiplet
can be conveniently denoted as $({\bf 1, 2, 1})\,$,
where the numerals count the numbers of the physical bosonic,
physical fermionic and auxiliary bosonic fields \cite{PT}. In this notation,
the previous ${\cal N}=1$ multiplet is $({\bf 1, 1, 0})$, and the ${\cal N}=2$ multiplet $({\bf 1, 2, 1})$ is split into
the direct sum of ${\cal N}=1$ multiplets as $({\bf 1, 2, 1}) = ({\bf 1, 1, 0}) \oplus ({\bf 0, 1, 1})\,$.}
 \be
\label{XM}
X^M = x^M + \theta \psi^M  +
 \bar \psi^M \bar \theta + F^M \theta \bar \theta \, .
 \ee
The action can be then written as \cite{FT}
  \be
\label{SN=2}
\!\!\!\!\!\!S &=& -\frac 12 \int dtd\bar\theta d\theta  \, g_{MN}(X) DX^M \bar D X^N \nn
\!\!\!\!\!\!&=& \frac 12 \int \, dt \Big[  g_{MN}\Big(\dot{x}^M \dot{x}^N
 + i[\bar\psi^M \nabla\psi^N -
\nabla\bar\psi^M \psi^N]\Big) +
R_{MNPQ} \bar \psi^M \psi^N \bar \psi^P \psi^Q \Big]
    \ee
with
 \be
D \ =\ \frac \partial {\partial \theta} - i \bar \theta \frac {\partial} {dt}\ , \ \ \ \ \ \ \ \ \ \ \ \ \
\bar D = - \frac  {\partial} {\partial \bar \theta} + i \theta \frac {\partial} {dt} \, .
 \ee
It involves a four-fermion term with the Riemann tensor. While passing in \p{SN=2} to the component action, we have eliminated the
auxiliary fields $F^M$ via their algebraic equations of motion.

It is well known that the system (\ref{SN=2}) has a nice
geometric interpretation \cite{Witgeom} :
the quantum supercharges can be interpreted as the exterior derivative operator
$d$ and its conjugate $d^\dagger 
$ of the de Rahm complex. 

The Lagrangians (\ref{SN=1}) and (\ref{SN=2}) can be written for an
arbitrary manifold. When the manifold is of some special type,
supersymmetric sigma models with further extended supersymmetries
can be defined. For instance, for a 3-dimensional manifold with
conformally flat metric and for the manifolds of dimension $3n$ with
metrics satisfying certain special conditions, the so-called
symplectic  ${\cal N} = 4$ supersymmetric sigma model can be defined
\cite{sympl}. For $5n$- dimensional manifolds with the metric
satisfying similar conditions supplemented by the harmonicity
conditions, one can write an interesting ${\cal N} = 8$ model
\cite{DESSI}. Actually, by now the whole ``zoo'' of ${\cal N} = 4$
and ${\cal N} = 8$ models is known, in both the manifestly
supersymmetric superfield off-shell formulations and the on-shell
component ones (see, e.g., \cite{ABC} and \cite{DI}, and references
therein). The general constraints on the target geometry required
for one or another type of extended supersymmetry were given, e.g.,
in \cite{PaCo}, \cite{GPS} and \cite{Hull}. The characteristic
feature of such geometries is that, in general, they involve
torsion, though torsionless geometries are admissible as well. In
particular, it is well known that, when the manifold is K\"ahler,
the Lagrangian (\ref{SN=2}) admits a second pair of supercharges
\cite{Zumino,Davis} and, when it is hyper-K\"ahler, three extra such
pairs exist \cite{HypK} (extending the supersymmetry  up to ${\cal N} = 4$
and  ${\cal N} = 8$, respectively). Also, the Lagrangian (\ref{SN=1}) for these
two types of the bosonic geometry admits an extension to the ${\cal N} = 2$ and ${\cal N} = 4$
supersymmetric ones.

In a recent paper \cite{IS}, a certain special
${\cal N}=2$ SQM sigma model for a
generic {\it complex} manifold of real dimension $D = 2n$ was
constructed and studied. One can introduce chiral superfields
   \be
Z{\,}^j = z{\,}^j +\sqrt{2}\,\theta \psi{\,}^j - i\theta\bar\theta \,\dot{z}{\,}^j\,,
\quad \bar Z{\,}^{\bar{j}} = \bar z{\,}^{\bar{j}}
-\sqrt{2}\,\bar\theta \bar\psi{\,}^{\bar{j}} +
i\theta\bar\theta \dot{\bar{z}}{\,}^{\bar{j}}\,,
   \ee
$j, \bar j = 1,\ldots,n$; $\bar D Z = D \bar Z = 0\,$, which describe ${\cal N}=2, d=1$ multiplets $({\bf 2, 2, 0})\,$.
The action was chosen in the form \cite{Hull}
\be
\lb{start}
&& S = \int dt d^2\theta \left({\cal L}_\sigma  + {\cal L}_{gauge}\right), \nn
&&
{\cal L}_\sigma = -\frac{1}{4}\,
h_{j\bar k}(Z, \bar Z)\, D Z{\,}^j \bar D\bar Z{\,}^{\bar k}\,, \qquad {\cal L}_{gauge} =
\, W(Z, \bar Z)\,
\ee
 with arbitrary superfunctions $ h_{i\bar j}$ (the metric, $ds^2 = 2 h_{j\bar k} dz^j d\bar z^{\bar k}$ ) and $W$
 (the prepotential from which the coupling to the background gauge potential
$A_M = ( -i\partial_j W, i\partial_{\bar j} W)$ is derived).
 The component action corresponding to \p{start} involves the torsion terms which disappear only for K\"ahler manifolds,
 when the metric satisfies the constraint $\partial_{[j}h_{i]\bar k} = 0\,$ (and its c.c.). The relevant quantum ${\cal N}=2$
 supercharges can be interpreted as the holomorphic exterior derivative $\partial$ and its conjugate $\partial^\dagger$,
 forming twisted or untwisted Dolbeault complexes.

Each of the actions (\ref{SN=1}), (\ref{SN=2}), and (\ref{start}) can be deformed to include extra torsions.
 Consider first
the action (\ref{SN=1}). It can be deformed by adding a term \cite{GPS}
 \be
\lb{torsionN=1}
   S \ =\ - \frac 1{12}\, \int d\theta dt\, C_{KLM} {\cal D}
{\cal X}^K {\cal D}{\cal X}^L {\cal D}{\cal X}^M \ .
  \ee
This gives the following component Lagrangian
  \be
\lb{LN1torscomp}
L =  \frac 12 \,  g_{MN}(x) \left( \dot{x}^M \dot{x}^N + i \psi^M \hat \nabla \psi^N \right)
 - \frac 1{12}\, \partial_P C_{KLM} \psi^P \psi^K \psi^L \psi^M \, ,
    \ee
where the covariant derivative $\hat \nabla$ involves now the {\it torsionful} affine connection
\be
\label{Gamhat}
\hat \Gamma_{K, LM} =
\Gamma_{K,LM} + \frac 12\, C_{KLM}\, ,
 \ee
 $\Gamma_{K, LM}$  being the standard Christoffel symbol.

The quantum supercharge
derived from the action (\ref{LN1torscomp}) with generic $g_{MN}, C_{KLM}$ has the form \cite{Braden2}
 \be
\label{Qreal}
{\cal Q}  \ =\ \psi^M  \left[ \Pi_M -  \frac i2\,  \Omega_{M, BC} \psi^B \psi^C \right ] +
\frac i{12}\, C_{KLM}  \psi^K \psi^L \psi^M \ ,
 \ee
where $\Omega_{M, BC}$ are the standard spin connections
satisfying the Cartan-Maurer equation
 \be
\lb{Cartan0}
de_A + \Omega_{AB} \wedge e_B = 0 \,, \quad {\rm whence}\quad \Omega_{M, AB} = e_{AK}(\partial_M e_B^K + \Gamma^K_{ML}e^L_B)\,.
 \ee

 The supercharge (\ref{Qreal}) can be interpreted as
a torsionful Dirac operator, where the torsions enter with an extra factor 1/3 \cite{Braden2,Mavra}. Indeed, the last term
in (\ref{Qreal}) can be absorbed into the following redefinition of the spin connection (cf. (\ref{Gamhat}))
$$
\Omega_{M, BC} \; \rightarrow \; \tilde{\Omega}_{M, BC} = e_{BK}(\partial_M e_B^K + \tilde{\Gamma}_{ML}^K e^L_C),\quad
\tilde\Gamma_{K, LM} =
\Gamma_{K,ML} + \frac 16\, C_{KML}\,.
$$

The action (\ref{SN=2}) for the  $({\bf 1,2,1})$  multiplet that corresponds to  de Rham complex can also be deformed.
In the geometrical language, the simplest such deformation \cite{Witgeom} is  described as
 \be
\label{deformW}
d_W {\cal O} & =& d {\cal O} -  dW \wedge {\cal O}\,, \nonumber \\[6pt]
d^\dagger_W {\cal O} & =& d^\dagger {\cal O} +
\langle dW , {\cal O} \rangle\,,
\ee
where $W$ is an arbitrary regular function and  $\langle dW , {\cal O} \rangle$   stands for
the interior product. For a $p$-form ${\cal O}$,
$$\langle dW , {\cal O} \rangle \ = \ p \,
 (\partial_{M_1} W)\, {\cal O}^{M_1}_{\ \ M_2\ldots M_p} \, dx^{M_2} \wedge \cdots \wedge  dx^{M_p} \, . $$
The deformation \p{deformW} corresponds to adding the potential term
 \be
\label{SW}
\int d^2\theta dt \, W(X^M)
 \ee
 to the action.

 One can consider also a deformation of a different type \cite{Braden1,RohmWit}
 \be
\label{deformdd+}
d_{\cal B} {\cal O} & =& d {\cal O}  -  d{\cal B} \wedge {\cal O}\,, \nonumber \\[6pt]
d^\dagger_{\cal B} {\cal O}  & =& d^\dagger {\cal O} -
\langle d \bar {\cal B} , {\cal O} \rangle\ ,
 \ee
where ${\cal B}$ is a regular 2-form (generically, complex) and
$\langle d\bar {\cal B} , {\cal O} \rangle$ is the interior product
involving the contraction of all the  indices in  $ d\bar{\cal B}\,$.
When ${\cal O}$ is a p-form with $p < 3$, it vanishes. The precise definition of
$\langle d\bar {\cal B} , {\cal O} \rangle  $ will be given in  \p{explicit} below.
 The exterior derivative of ${\cal B}$ can be associated
with the torsion  ${\cal C}$.

 One should note here that, in contrast to the systems with  $({\bf 1,1,0})$ or
 $({\bf 2,2,0})$ multiplets, the  analogy between $d  {\cal B}$ and the torsion  in the case of the
 $({\bf 1,2,1})$  multiplet is not quite direct.
First of all, the torsions are usually assumed real, while  ${\cal B}$ and  $d  {\cal B}$ can be complex.
Second, as we will see in Sect. 3, in this case, the torsions enter covariant derivatives not in the same way
as standard Christoffel symbols: $\sim {\cal C}\psi\psi$ vs. $\sim \Gamma \bar\psi \psi$.
All this notwithstanding, we will call
 $ {\cal C}$ torsion also in this case.

One can easily observe that the operators
$  d_{\cal B}, d^\dagger_{\cal B}$ (as well as the operators $d_W, d^\dagger_W$)
 are nilpotent. They form thus
a minimal supersymmetry
algebra by the same token as the operators $d, d^\dagger$ do. This deformed supersymmetry system can be realized
in the superfield language. To this end, one should add the term
 \be
\label{realB2}
S_2 =  \frac{1}{2} \int d^2 \theta  dt \, {\cal B}_{MN}(X^P)\, DX^M DX^N  \ + \ {\rm c.c.}
 \ee
to the action \p{SN=2} \cite{GPS}. The expressions for the quantum supercharges
and the Hamiltonian for this model (in the case of real ${\cal B}_{MN}$) can be found in \cite{Kimura}.

One can deform the system (\ref{SN=2}) even further by adding the exterior derivative
$d {\cal B}_4$ of an arbitrary 4-form ${\cal B}_4$ to $d_{\cal B}$.
The deformed operator
$d_{W, {\cal B}_{2}, {\cal B}_{4}}$ is still
nilpotent. In superfield language, that corresponds to adding
the structure
    \be
\label{realB4}
S_4 \ =  \frac{1}{2}\int d^2 \theta dt \, {\cal B}_{MNPQ}(X^S)\, DX^M DX^N DX^P DX^Q  \ + \ {\rm c.c.}
 \ee
to the action. The component Lagrangian of a  model with
an extra 4-form will be written in Sect. 2 below.
One can further add the exterior derivative of a 6-form, etc. These higher even-dimensional forms
can be dubbed {\it generalized} torsions. It should be pointed out that these additional superfield terms do not
bring in the component Lagrangians any terms of higher order in time derivatives of the involved fields.

The complex sigma model Lagrangian (\ref{start}) can be generalized along similar lines.
 The generic Lagrangian is obtained by adding
the terms
   \be
\label{Bterms}
   \sim {\cal B}_{jk}(Z, \bar Z)D Z{\,}^{j} D Z{\,}^{k} + {\rm c.c.}\,
   \ee
to ${\cal L}_\sigma\,$\cite{Hull}. By the same token as in the sigma model involving real $({\bf 1,2,1})$
multiplets, one can also add the terms $\propto {\cal B}_{jklp}$, etc.
The  terms, associated with extra torsions (torsions coming from (\ref{Bterms}) should be added to
the torsions which are already present in (\ref{start}) in non-K\"ahler case),
as well as with the generalized torsions, were not  considered in \cite{IS}. The corresponding
supercharges define some ${\cal B}$-deformation of the Dolbeault complex.

The present paper is devoted to filling  some gaps existing in the literature
on these subjects.
In particular, we give the explicit form of the ${\cal N}{=}2$ supercharges for the
torsionful sigma models based on the multiplets $({\bf 1,2,1})$ and
$({\bf 2,2,0})$, with taking into account, in the first case, both
interactions \p{realB2} and \p{realB4}, as well as the potential
term \p{SW}.
We also find the vacuum states in these sigma models and demonstrate
that the inclusion of the torsion terms does not influence their number.

The plan of the paper is the following.
In Sect. 2, we discuss the $({\bf 1,2,1})$ multiplet. We write the Lagrangian and
present  both the classical and the quantum supercharges in the system that includes the torsions
and generalized torsions.

In Sect. 3, we discuss the complex sigma model. We write a generic component Lagrangian, the supercharges
and the Hamiltonian, both at the classical and the quantum levels for the model involving the terms (\ref{Bterms}).

  In Sect. 4, we are addressing the supersymmetric vacua of our models. There is a simple mathematical argument
saying that cohomology classes of the deformed de Rham complex \p{deformW}, \p{deformdd+} are the same as for the undeformed one.
A similar reasoning applies to the ${\cal B}$-deformed Dolbeault complex too.
To illustrate and confirm these statements,
 we present some explicit calculations for the wave functions of deformed vacuum states on the spheres $S^n$
 for the real sigma model and on the $\mathbb{CP}^n$ manifolds for the complex one.

\setcounter{equation}{0}
\section{Torsions and generalized torsions for $({\bf 1,2,1})$ sigma model}

Consider the supermultiplet (\ref{XM}).
${\cal N}=2$ supersymmetry acts there as
\be
\delta X^M = -(\epsilon Q + \bar\epsilon \bar Q)X^M\,, \qquad Q = \frac{\partial}{\partial \theta} +
i\bar\theta\partial_t\,,
\quad \bar Q = \frac{\partial}{\partial \bar\theta} + i\theta\partial_t\,, \lb{tranZA}
\ee
whence we obtain the transformations of the component fields
\be
&& \delta x^M = - (\epsilon\,\psi^M - \bar\epsilon \bar\psi^M)\,, \nn
&& \delta \psi^M = \bar\epsilon\,(i\dot{x}^M - F^M)\,, \quad
\delta \bar\psi^M = -\epsilon\,(i\dot{x}^M + F^M)\,, \nn
&& \delta F^M =i(\epsilon\, \dot\psi{}^M + \bar\epsilon\,\dot{\bar\psi}{}^M)\,.\lb{N22}
\ee

 Write the action as the sum
$S = S_g + S_2 + S_4$, where
$S_g$ is the standard action given by the sum of (\ref{SN=2}) and (\ref{SW}), while $S_2$ and $S_4$ are the terms
(\ref{realB2}) and (\ref{realB4}) describing the torsions and generalized torsions.

The corresponding component actions  have the form
\be
S_g &=& \int dt\Big[ \frac{1}{2}\,g_{MN}\left(\dot{x}^M \dot{x}^N + F^MF^N\right)
+\frac{i}{2}\, g_{MN}\left(\bar\psi^M \nabla\psi^N -
\nabla\bar\psi^M \psi^N \right) \nn
&&+\,\Gamma_{M, NP}\psi^N\bar\psi^PF^M- \frac 12\, \partial_{[M}\Gamma_{N, P]Q}\psi^M\bar\psi^Q\psi^P\bar\psi^N \nn
&&+\, F^M\partial_MW + \partial_M\partial_N W\psi^M\bar\psi^N\Big], \lb{Sh}\\
S_2 &=& \frac{1}{2}\int dt\Big[\left(\partial_M C_{NPQ}\psi^N\psi^P\psi^Q\bar\psi^M +
\partial_M \bar{C}_{NPQ}\bar\psi^N\bar\psi^P\bar\psi^Q\psi^M\right) \nn
&&+\,\, 3 \,C_{MNP}(F^M -i\dot{x}^M)\psi^N\psi^P
-  3\,\bar{C}_{MNP}(F^M + i\dot{x}^M)\bar\psi^N\bar\psi^P\Big], \lb{S22} \\
S_4 &=& \frac{1}{2}\int dt \Big[\left(\partial_M C_{NPQST}\psi^N\psi^P\psi^Q\psi^S\psi^T\bar\psi^M + {\rm c.c.}\right) \nn
&&+\,\,5\, C_{MNPQS}(F^M - i\dot{x}^M)\psi^N\psi^P\psi^Q\psi^S +{\rm c.c.}\Big]. \lb{S42}
\ee
Here
\be
&& \nabla\psi^M = \dot\psi^M + \Gamma^M_{N P}\dot{x}^N\psi^P\,, \quad \Gamma_{M, NP} =
\frac{1}{2}\left[\partial_N g_{MP} + \partial_P g_{MN} - \partial_M g_{NP} \right], \nn
&&C_{MNP} = \frac{1}{3}\left\{\partial_M {\cal B}_{NP} + {\rm cycle} (M, N, P)\right\}, \nn
&&C_{MNPQS} = \frac{1}{5}\left\{\partial_M {\cal B}_{NPQS} + {\rm cycle} (M, N, P, Q, S)\right\}.
\ee

The $F^M$ equation of motion (with $W=0$ for  simplicity) yields:
\be
F^M = -\Gamma^M_{NP}\,\psi^N\bar\psi^P -  \frac{1}{2}\,{\cal C}^M\,,
\ee
where
\be
{\cal C}^M = 3\left[ C^{(2)M}(\psi)^2 - \bar{C}^{(2)M}(\bar\psi)^2\right] + 5 \left[ C^{(4)M}(\psi)^4
+ \bar{C}^{(4)M}(\bar\psi)^4\right],
\ee
and we used the condensed notation
\be
C^{(2)M}(\psi)^2 := C^M_{\;\;NP}\,\psi^N\psi^P\,, \quad C^{(4)M}(\psi)^4 :=C^M_{\;\;NPQS}\,\psi^N\psi^P\psi^Q\psi^S\,,
\quad {\rm etc}.
\ee

After substitution of this expression back into the sum of actions \p{Sh} - \p{S42} (with $W=0$)  we obtain the on-shell
form of the total action
\be
\!\!\!\!\!\!S_{\rm on-sh} &=& \frac 12\int dt \Big[g_{MN}\dot{x}^M \dot{x}^N  + i g_{MN}(\bar\psi^M \nabla\psi^N -
\nabla\bar\psi^M \psi^N) + R_{MNPQ}\bar\psi^M \psi^N \bar\psi^P\psi^Q \nn[6pt]
\!\!\!\!\!\!&&+\,(\nabla_M C_{NPQ}\psi^N\psi^P\psi^Q\bar\psi^M +
\nabla_M \bar{C}_{NPQ}\bar\psi^N\bar\psi^P\bar\psi^Q\psi^M) \nn[6pt]
\!\!\!\!\!\!&& -\,3i\,\dot{x}^M(C_{MNP}\psi^N\psi^P + \bar{C}_{MNP}\bar\psi^N\bar\psi^P) \nn[6pt]
\!\!\!\!\!\!&& + \,(\nabla_M C_{NPQST}\psi^N\psi^P\psi^Q\psi^S\psi^T\bar\psi^M
-\nabla_M \bar{C}_{NPQST}\bar\psi^N\bar\psi^P\bar\psi^Q\bar\psi^S\bar\psi^T\psi^M) \nn[6pt]
\!\!\!\!\!\!&& -\,5i\,\dot{x}^M(C_{MNPQS}\psi^N\psi^P\psi^Q\psi^S
- \bar{C}_{MNPQS}\bar\psi^N\bar\psi^P\bar\psi^Q\bar\psi^S) - \frac 1{4}\, {\cal C}^M{\cal C}_M \Big], \lb{onshellS}
\ee
where
\be
R_{MNPQ} = g_{MT}\left(\partial_P\Gamma^T_{QN} -  \partial_Q\Gamma^T_{PN}  + \Gamma^T_{PS}\Gamma^S_{QN}
-\Gamma^T_{QS}\Gamma^S_{PN}
\right)
\ee
is the Riemann tensor.

When deriving the supercharges, it is more convenient not to eliminate the auxiliary field $F^M$ until the final step.
The classical conserved Noether supercharge calculated from the infinitesimal transformations \p{N22} that leave invariant the
off-shell action $S = S_g + S_2 + S_4 $  has the following form,
\be
Q &=& \psi^M (\tilde \Pi_M + i\partial_M W) -\frac{i}{2}\,\partial_M g_{NP}\psi^M\psi^N\bar\psi^P \nn
&&+ \,i\, C_{MNP}\psi^M\psi^N\psi^P +  \, i   C_{MNPQS}\psi^M\psi^N\psi^P\psi^Q\psi^S\,, \lb{Q}
\ee
where $\tilde \Pi_M$ is the canonical momentum,
\be
&&\tilde \Pi_M = \frac{\partial L}{\partial \dot{x}^M} = g_{MN}\dot{x}^N
- \frac{i}{2}\,\Gamma_{N, MP}\left(\psi^P\bar\psi^N - \psi^N\bar\psi^P\right) - \frac{i}{2}\,{\cal S}_M\,, \lb{Pi}
 \ee
with
$${\cal S}^M :=  3 \left[ C^{(2)M}(\psi)^2 + \bar{C}^{(2)M}(\bar\psi)^2\right] + 5 \left[ C^{(4)M}(\psi)^4
- \bar{C}^{(4)M}(\bar\psi)^4\right]. $$

Correspondingly,
\be
\bar Q &=& \bar\psi^M (\tilde{\Pi}_M - i\partial_M W) -\frac{i}{2}\,\partial_M h_{NP}\bar\psi^M\bar\psi^N\psi^P \nn
&&+ \,i  \, \bar{C}_{MNP}\bar\psi^M\bar\psi^N\bar\psi^P
- i\,  \bar{C}_{MNPQS}\bar\psi^M\bar\psi^N\bar\psi^P\bar\psi^Q\bar\psi^S\,.\lb{barQ}
\ee
 Even though the auxiliary fields $F^M$ were kept in the Lagrangian, they do not explicitly appear in the supercharges.

The partial derivative (\ref{Pi}) was calculated assuming fixed $\psi^M, \bar \psi^M$. The latter variables are not, however,
 canonically conjugated, their Poisson bracket being $\{\bar \psi^M, \psi^N\} = g^{MN}$. As a preliminary step to quantization,
one should define the tangent space canonically conjugated fermion variables
$\psi^A = e^A_M \psi^M$, $\bar\psi^A = e^A_M \bar\psi^M$ and express the classical
supercharges through these variables and the new bosonic canonical momentum
  \be
\Pi_M =  \left. \frac{\partial L}{\partial \dot{x}^M} \right|_{{\rm fixed} \ \psi^A}
= \tilde \Pi_M
- \frac{\partial \dot{\psi}{}^A}{\partial \dot{x}{}^M}\,\frac {\partial L}{\partial \dot{\psi}{}^A} -
 \frac{\partial \dot{\bar\psi}{}^A}{\partial \dot{x}{}^M}\,\frac {\partial L}{\partial \dot{\bar\psi}{}^A}\, .
\ee
Finally, we derive the following classical expression for $Q$ (and analogously for $\bar Q$):
\be
Q &=& \psi^A\,e^{AM} (\Pi_M + i\partial_M W) - i\Omega^{C,AB}\,\psi^C\psi^A\bar\psi^B \nn
&&+ \,i\,C^{ABC}\psi^A\psi^B\psi^C + i\,  C^{ABCDE}\psi^A\psi^B\psi^C\psi^D\psi^E\,. \lb{Q2}
\ee
Here,
\be
\Omega^{C,AB} = e^{MC}\,\Omega_{M}^{AB}\,, \quad \Omega_{M}^{AB} = e^A_N(\partial_M e^{N B} + \Gamma^N_{MT}e^{TB})\,,
\ee
is the standard spin connection.

To derive the  quantum supercharges, one has to replace $\Pi_M$ and $\bar \psi^A$ by differential
operators $\Pi_M \to -i \partial/\partial x^M$, \ $\bar \psi^A  \to  \partial /\partial \psi^A$ and to resolve the ordering
ambiguities problem. To make a selection between many different quantum theories corresponding to a given
classical one,  we require that the supersymmetry algebra
remains intact at the quantum level and that $Q_{qu}$ and $\bar Q_{qu}$ are Hermitian conjugate
to each other. This fixes the quantum supercharges and Hamiltonian.

As was shown in \cite{howto}, the general recipe of such a symmetry-preserving quantization is as follows.
 \begin{itemize}
 \item Take the expressions for the classical supercharges and order them according to the symmetric Weyl prescription.
 \item The  supercharges thus obtained are nilpotent. Their anticommutator  gives  the quantum Hamiltonian. Generically, it does
{\it not} coincide with the operator obtained from the classical Hamiltonian by Weyl ordering.
 \item This procedure gives the quantum supercharges and the Hamiltonian acting in the ``flat'' Hilbert space
with the measure
$$ \sim \left( \prod_M dx^M \right) d({\rm fermions}) $$
in the inner product. If we want to obtain the expressions for the {\it covariant}
operators acting in the Hilbert space
with the measure involving the factor $\sqrt{\det g}$ (such operators have a nicer geometric interpretation),
an appropriate similarity transformation
 \be
\lb{similar}
(\,Q^{\rm cov}, \bar Q^{\rm cov}\,) = (\det g)^{-1/4}\,(\,Q^{\rm flat}, \bar Q^{\rm flat}\,)(\det g)^{1/4}\,
\ee
should be performed.
  \end{itemize}

We finally obtain the quantum supercharges as
  \be\lb{Q121-cov}
Q^{\rm cov} &=&  -i\psi^Ae^{MA}(\partial_M - \partial_M W) -i  \Omega^{C, AB}\,\psi^C\bar\psi^A\psi^B \nn[4pt]
&&\qquad+ \,i\,  C^{ABC}\psi^A\psi^B\psi^C + i\, C^{ABCDE}\psi^A\psi^B\psi^C\psi^D\psi^E\,, \nn[6pt]
\bar Q^{\rm cov} &=& -i \bar\psi^Ae^{MA}(\partial_M + \partial_M W)-i \Omega^{C, AB}\,\bar\psi^C\psi^A\bar\psi^B \nn[4pt]
&&\qquad+ \,i \, \bar{C}{}^{ABC}\bar\psi^A\bar\psi^B\bar\psi^C - i\,
\bar{C}{}^{ABCDE}\bar\psi^A\bar\psi^B\bar\psi^C\bar\psi^D\bar\psi^E\,.
\ee

 The torsion-free part of these expressions is well known \cite{Davis,Claudson}. The terms involving the torsions
$C_{ABC}$ were written in Ref. \cite{Kimura} (for real $C_{ABC}$). The supercharges thus obtained
constitute the ${\cal N}{=}2, d=1$ Poincar\'e superalgebra
\be
{\rm (a)} \; (Q^{\rm cov})^2 = (\bar Q^{\rm cov})^2 =0\,, \; {\rm (b)}\; \{Q^{\rm cov}, \bar Q{}^{\rm cov}\} = 2 H\,, \;
{\rm (c)}\, [Q^{\rm cov}, H] = [\bar Q{}^{\rm cov}, H] = 0\ \lb{N2poinc}
\ee
and are isomorphic to the twisted de Rham operators  (\ref{deformdd+}) with ${\cal B} = {\cal B}_{2}
+ {\cal B}_{4}$. For our further purposes, we will not need the explicit form of the quantum Hamiltonian $H\,$.
For real $C^{ABC}$ and vanishing $C^{ABCDE}$, it was given in \cite{Kimura}. The isomorphism between the
supercharges \p{Q121-cov} and the operators appearing in the geometric setting (\ref{deformdd+}) implies, in particular, the
correspondence
\be
\langle d \bar {\cal B} , {\cal O} \rangle\;
&\Leftrightarrow & \;\bar C^{ABC}\frac{\partial}{\partial \psi^A}\frac{\partial}{\partial \psi^B}
\frac{\partial}{\partial \psi^C}\,\psi^{D_1}\ldots \psi^{D_n}\,{\cal O}^{D_1 \ldots D_n} \,,
\lb{correspond}
\ee
which gives the
following explicit definition for the inner product in the second line of (\ref{deformdd+}):
\be
\langle d \bar {\cal B} , {\cal O} \rangle \ =\
 -\frac{p!}{(p-3)!}\,
\bar C^{MNP}{\cal O}_{MNP R_4 \ldots R_n}
dx^{R_4} \wedge \cdots \wedge dx^{R_p}
\lb{explicit}
\ee
for a $p$-form ${\cal O}$ ($p \geq 3$).
We will use it in what follows.

The supercharges \p{Q121-cov} can be represented in the form
\be
Q^{\rm cov} \!\!\!&=&\!\!\!
e^{W+ \psi^K\psi^L\mathcal{B}_{KL}+\psi^K\psi^L\psi^M\psi^N\mathcal{B}_{KLMN}} \,Q^{\rm cov}_{0} \,
e^{-W-\psi^K\psi^L\mathcal{B}_{KL}-\psi^K\psi^L\psi^M\psi^N\mathcal{B}_{KLMN}}\,,\nn[6pt]
\bar{Q}^{\rm cov} \!\!\!&=&\!\!\!
e^{-W + \bar\psi{}^K\bar\psi{}^L\bar{\mathcal{B}}_{KL}
- \bar\psi{}^K\bar\psi{}^L\bar\psi{}^M\bar\psi{}^N\bar{\mathcal{B}}_{KLMN}} \,
\bar{Q}^{\rm cov}_{0} \,
e^{W - \bar\psi{}^K\bar\psi{}^L\bar{\mathcal{B}}_{KL}
+\bar\psi{}^K\bar\psi{}^L\bar\psi{}^M\bar\psi{}^N\bar{\mathcal{B}}_{KLMN}},\label{Qcov-121}
\ee
where $Q^{\rm cov}_{0}, \bar{Q}^{\rm cov}_{0}$ are the supercharges  with the torsion and potential terms
being suppressed. Note that the derivatives in   $Q^{\rm cov}_{0}, \bar{Q}^{\rm cov}_{0}$ act here not only
on $W$ and ${\cal B}$, but also
on $\psi^K = e^{K A} \psi^A$, etc. These terms are exactly canceled by the terms coming from the commutators of the structures
$\sim \Omega \psi \bar\psi \psi $ and   $\sim \Omega \bar \psi \psi \bar \psi $
in   $Q^{\rm cov}_{0}$ and $ \bar{Q}^{\rm cov}_{0}$
with the torsion structures.

In the differential form language, this notable representation of the supercharges
has a rather transparent meaning. The first line in \p{Qcov-121} means that
 \be
\lb{ddefexp}
 d_{W, {\cal B}} = e^{W + \cal B} d e^{-W - \cal B} \, ,
 \ee
which is a direct corollary of the definitions \p{deformW}, \p{deformdd+}.
The second line is Hermitian conjugate of the first one.
The representation \p{Qcov-121}, \p{ddefexp}  will be used
 while finding the explicit form of the ground
state wave functions in Section 4.

\setcounter{equation}{0}

\section{Complex model with torsions}

We start from the action (\ref{start}) and add to it the term
 \be
\lb{extra}
 S_{\rm extra\ torsion} = \frac 14 \int dt d^2\theta
\left(  {\cal B}_{jk}(Z, \bar Z)\,D Z{\,}^{j} D Z{\,}^{k} -
\bar{\cal B}_{\bar j\bar k}(Z, \bar Z)\,\bar D
\bar Z{\,}^{\bar j} \bar D \bar Z{\,}^{\bar k}
  \right)
\ee
with arbitrary antisymmetric complex superfunction ${\cal B}_{jk}$ and its conjugate
$\bar {\cal B}_{\bar j\bar k}\,$.

The component form of the full action is
\be
S
&=& \int dt \Big\{ h_{j\bar k} \left[\dot{z}{\,}^j\dot{\bar z}{\,}^{\bar k}
+\frac{i}{2} \left( \psi{\,}^j \dot{\bar{\psi}}{\,}^{\bar k} - \dot\psi{\,}^j \bar{\psi}{\,}^{\bar k}\right)\right]
+ \left(\partial_t\partial_{\bar l} h_{j\bar k}\right) \psi{\,}^t\psi{\,}^j \bar\psi{\,}^{\bar l}\bar\psi{\,}^{\bar k}\nn
&& \qquad -\, \frac{i}{2}\left[\left(2\partial_j h_{t\bar k} - \partial_t h_{j\bar k}\right)\dot z{\,}^t
-  \left(2\partial_{\bar k} h_{j \bar t} -
\partial_{\bar t} h_{j\bar k}\right)\dot{\bar z}{\,}^{\bar t}\right]\psi{\,}^j \bar\psi{\, }^{\bar k} \nn
&& \qquad +\, 2 \partial_j\partial_{\bar k}W\,\psi{\,}^j\bar\psi{\,}^{\bar k} -  i \left(\partial_j W\dot z{\,}^j
- \partial_{\bar j}W \dot{\bar z}{\,}^{\bar j}\right)   \nn
&& \qquad -\,3i\,\partial_{[m}{\cal B}_{ik]}\,\dot z{\,}^{m} \psi{\,}^{i} \psi{\,}^{k}
-3i\,\partial_{[\bar m}\bar {\cal B}_{\bar i\bar k]}\,
\dot {\bar z}{\,}^{\bar m} \bar \psi{\,}^{\bar i} \bar \psi{\,}^{\bar k}   \nn
&& \qquad -\,\partial_{\bar n}\partial_{m}{\cal B}_{ik}\,\bar\psi{\,}^{\bar n} \psi{\,}^{m} \psi{\,}^{i} \psi{\,}^{k}
- \partial_{n}\partial_{\bar m}\bar {\cal B}_{\bar i\bar k}\,
\psi{\,}^{n}\bar \psi{\,}^{\bar m} \bar \psi{\,}^{\bar i} \bar \psi{\,}^{\bar k} \Big\}\,. \lb{genAct}
\ee

This action can be cast in the ${\cal N} =1$ superfield notations as the sum of the terms
(\ref{SN=1}), (\ref{torsionN=1}) and the term
  \be
\lb{gaugeN=1}
S_{\rm gauge} \ =\ -i  \int dt d\theta \,  A_M({\cal X}^P)  {\cal D} {\cal X}^M \ ,
 \ee
with  $M = \{j, \bar j\}$, $A_M = \{-i \partial_j W, i \partial_{\bar j} W \}\,$.
 This gives in components
\be
L &=& L_\sigma + L_{\rm extra\,\,torsion} + L_{\rm  gauge}\nn
 &=& \frac{1}{2}\left[ g_{MN}\,\dot z{\,}^M \dot z{\,}^N + ig_{MN}\,\psi^M \hat \nabla \psi^N
- \frac{1}6 \,\partial_P C_{KLM}\,\psi^P\psi^K\psi^L\psi^M \right] \nn
&& +\, A_M \dot{z}^M  - \frac i2\, F_{MN} \psi^M \psi^N\, ,  \lb{1comp}
\ee
where $F_{MN} = \partial_M A_N - \partial_N A_M\,$ and
$z^M \equiv x^M = (z^j, \bar z^{\bar j}), \psi^M =(\psi^j, \bar \psi^{\bar j})\,$. It is worth pointing out that the target
space in this case is even-dimensional.

The non-vanishing components of the totally antisymmetric torsion tensor $C_{KLM}$ are
\be
\label{Cikl}
C_{k l\bar m} = -\left(\partial_{k} h_{l\bar m} - \partial_{l} h_{k\bar m}\right) \,,  &\quad&
C_{\bar k \bar l  m} =  (C_{k l \bar m})^* = -\left(\partial_{\bar k} h_{m \bar l} - \partial_{\bar l} h_{m \bar k}\right)\,,
\nn [5pt]
C_{k l m} = 12\, \partial_{[k} \mathcal{B}_{lm]}  \,,  &\quad&
C_{\bar k \bar l  \bar m} =  12\, \partial_{[\bar k} \bar {\mathcal{B}}_{\bar l\bar m]}\ .
\ee

 We see that the terms $\propto {\cal B}_{jk}, \, \bar {\cal B}_{\bar j\bar k}$ in the Lagrangian bring about the holomorphic
components of the torsion $C_{klm}, C_{\bar k \bar l \bar l}$. The (3,0)-form $C_{klm} dz^k \wedge dz^l \wedge
dz^m$ is obtained from the arbitrary (2,0)-form ${\cal B}_{jk} dz^j \wedge dz^k$ by the action of the exterior holomorphic
derivative $\partial$.
Besides, there are mixed components of the torsion
tensor $C_{k l\bar m}, C_{\bar k \bar l  m}$ which are not arbitrary, but are strictly related to the metric
$h_{j\bar k}$. When the manifold is K\"ahler, i.e., $h_{j \bar k} = \partial_j \partial_{\bar k} K \,$,
these components vanish. In this case (and when ${\cal B}_{jk} = 0$), the Lagrangian coincides with (\ref{SN=1}).

Generically, the Lagrangian in Eqs. (\ref{genAct}), (\ref{1comp}) involves a 4-fermion term.
Note that, if the form $C_{MNK} dz^M \wedge dz^N \wedge dz^K$ is closed, the 4-fermion term is absent
(this case was addressed in \cite{Mavra}). Note also that the ``new'' terms brought about by the extra
torsion terms $\propto  {\cal B}, \bar {\cal B}$ show up only starting from the complex target dimension $n=3$.  For $n=2$
(and, of course, for $n=1$) they vanish identically.

The classical supercharges can be calculated by the N\"other theorem in a standard way.
We obtain
\begin{equation}\label{Qclass}
\begin{array}{rcl}
Q &=& \sqrt{2}\,e^k_c\psi{\,}^c\left[ \Pi_k - i \,\Omega_{k, \bar a b}\,\bar\psi{\,}^{\bar a} \psi{\,}^b
+ i\psi{\,}^{a}  \psi{\,}^b e^j_a e^l_b \partial_k {\cal B}_{jl} \right] , \\[7pt]
\bar Q&=& \sqrt{2}\,e^{\bar k}_{\bar c}\bar\psi{\,}^{\bar c}\left[ \bar\Pi_{\bar k}  -
i \,\bar{\Omega}_{\bar k, a \bar b}\, \psi{\,}^{a} \bar\psi{\,}^{\bar b} +
i\bar \psi{\,}^{\bar a}  \bar \psi{\,}^{\bar b} e^{\bar j}_{\bar a} e^{\bar l}_{\bar b}
 \partial_{\bar k} {\bar {\cal B}}_{\bar j\bar l} \right],
\end{array}
\end{equation}
 where
\be
\Pi_k = P_{k}   + i \, \partial_k W\,, \qquad \bar\Pi_{\bar k} =
P_{\bar k} - i \,\partial_{\bar k} W\,,
  \ee
and $P_k, P_{\bar k}$ are the canonical momenta (obtained by varying the Lagrangian with respect to
$\dot{z}^k, \dot{\bar z}^{\bar k}$ at fixed $\psi^a, \bar\psi^{\bar a}$). The spin connections $\Omega_{k, \bar a b}$ and
$\bar \Omega_{\bar k, a \bar b}$ are the corresponding components of the {\it standard} real spin connections
$\Omega_{M, AB}$ satisfying (\ref{Cartan0}). The terms
$\propto \partial {\cal B}$ in $Q, \bar Q$ can be interpreted as
the holomorphic components
$\hat \Omega_{k, ab}$ and
$\hat {\bar \Omega}_{\bar k, \bar a \bar b}$ of the connection
 \be
\lb{hatOmega}
\hat \Omega_{M, AB} = e_{AN} (\partial_M e^N_B + \hat\Gamma^N_{MK} e^K_B) = \Omega_{M, AB} + \frac{1}{2}\,e_A^K e_B^L C_{MLK}\,,
 \ee
in which the torsion is taken into account and which satisfies the following generalization of \p{Cartan0}
  \be
\lb{Cartantor}
de_A + \hat \Omega_{AB} \wedge e_B = C_A = C_{ABC} \, dz^B \wedge dz^C\,.
 \ee

One can be convinced that,  for the {\it particular} torsion whose components are displayed in Eq.(\ref{Cikl}),
the sum $Q + \bar Q$ of the supercharges (\ref{Qclass}) coincides with the  $({\bf 1,2,1})$  supercharge
(\ref{Qreal}). Note that for a generic complex manifold, the spin connections involve the components
$\Omega_{k, \bar a \bar b}$ and $\Omega_{\bar k, ab}$ which vanish in the K\"ahler case. One can observe
that their contribution to the first term of Eq. (\ref{Qreal})
exactly cancels out the contributions due to $C_{jk\bar l}, C_{\bar j \bar k l}$
in the second term.

The canonical classical Hamiltonian  $H_{cl}$ can be represented in the following compact form:
\be
 \lb{Hclass}
 H_{cl} &=& h^{\bar k j}
 {\cal P}_{j}  \bar {\cal P}_{\bar k} -
e^t_a e^j_c e^{\bar l}_{\bar b} e^{\bar k}_{\bar d}\,(\partial_t\partial_{\bar l}{\,}h_{j\bar k})\,\psi{\,}^a\psi{\,}^c
\bar\psi{\,}^{\bar b}\bar\psi{\,}^{\bar d} \nn[6pt]
&& +\,e^{\bar m}_{\bar d}e^i_a e^j_c e^{k}_{c} \,
(\partial_{\bar m}\partial_{i}{\cal B}_{jk})\,\bar\psi{\,}^{\bar d} \psi{\,}^{a} \psi{\,}^{b} \psi{\,}^{c}
+ e^m_d e^{\bar i}_{\bar a}e^{\bar j}_{\bar b}e^{\bar k}_{\bar c}\,
(\partial_{m}\partial_{\bar i}\bar {\cal B}_{\bar j\bar k})\,
\psi{\,}^{d}\bar \psi{\,}^{\bar a} \bar \psi{\,}^{\bar b} \bar \psi{\,}^{\bar c}\,  \nn[6pt]
&& -\, 2\, e^j_a e^{\bar k}_{\bar b}\,(\partial_j\partial_{\bar k} W)\, \psi{\,}^a\bar\psi{\,}^{\bar b} \,,
\ee
where
 \be
{\cal P}_M = \Pi_M - \frac i2 \hat \Omega_{M, AB} \psi^A \psi^B \ .
 \lb{calP}
\ee
In contrast to the supercharge (\ref{Qreal}),  the Hamiltonian (\ref{Hclass}) involves
 the conventional ``hatted'' spin connections
(\ref{hatOmega}).

Let us now turn to quantum theory. As in the previous section, we resolve the ordering ambiguities as prescribed in
\cite{howto}, i.e. use the symmetric Weyl ordering for the supercharges supplemented by a similarity transformation
(\ref{similar}).
We obtain the following expressions for the covariant quantum supercharges:
 \be
\label{Qcovgen}
\begin{array}{rcl}
Q^{\rm cov} &=& \sqrt{2}\,e^k_c\psi{\,}^c\left[ \Pi_k - \frac i2 \partial_k (\ln \det \bar e) +
i \,\Omega_{k, \bar a b}\,\psi{\,}^b \bar\psi{\,}^{\bar a}
+ i\psi{\,}^{a}  \psi{\,}^b e^j_a e^l_b \partial_k {\cal B}_{jl} \right] , \\[7pt]
\bar Q^{\rm cov}&=& \sqrt{2}\,e^{\bar k}_{\bar c}\bar\psi{\,}^{\bar c}\left[ \bar\Pi_{\bar k}  -
\frac i2 \partial_{\bar k} (\ln \det e) +
i \,\bar{\Omega}_{\bar k, a \bar b}\, \bar\psi{\,}^{\bar b} \psi{\,}^{a}  +
i\bar \psi{\,}^{\bar a}  \bar \psi{\,}^{\bar b} e^{\bar j}_{\bar a} e^{\bar l}_{\bar b}
 \partial_{\bar k} {\bar {\cal B}}_{\bar j\bar l} \right].
\end{array}
\ee
These expressions almost coincide by form with (\ref{Qclass}) (note, however, the presence of important
terms $\propto \partial \,\ln \det e\,, \,  \bar \partial \,\ln \det \bar e\,$), but $\bar \psi^{\bar a}$ are now operators,
$\bar \psi^{\bar a} = \partial/\partial \psi^a\,$, and, similarly, $\Pi_M = -i\partial_M -A_M\,$, with
$A_M = \{-i \partial_j W, i \partial_{\bar j} W \}\,$.

 The quantum Hamiltonian is
  \be
\label{Hcovgen}
H^{\rm cov}_{qu} &=& - \frac 12\, \triangle^{\rm cov} + \ \frac 18 \left (R -
\frac 12\, h^{\bar i i}h^{\bar j j}h^{\bar k k}  C_{i\,j\, \bar k}\,C_{\bar i\,\bar j\, k} -
\frac{1}{6}\, h^{\bar ii}h^{\bar jj}h^{\bar kk}  C_{i\,j\, k}\,C_{\bar i\,\bar j\, \bar k} \right) \nn
&& -\, 2 \langle \psi^a \bar \psi^{\bar b} \rangle\,e^i_a e^{\bar j}_{\bar b}\,(\partial_i\partial_{\bar j} W)
- \langle \psi^a \psi^b \bar \psi^{\bar c} \bar \psi^{\bar d} \rangle \,
 e^i_a e^j_b e^{\bar k}_{\bar c} e^{\bar l}_{\bar d} \, (\partial_i\partial_{\bar k}{\,}h_{j\bar l})  \lb{kvantH}
\\[3pt] \nonumber
&& +\,\langle\bar\psi{\,}^{\bar d} \psi{\,}^{a} \psi{\,}^{b}
\psi{\,}^{c}\rangle\,e^{\bar m}_{\bar d}e^i_a e^j_c e^{k}_{c} \,
(\partial_{\bar m}\partial_{i}{\cal B}_{jk})
+  e^m_d e^{\bar i}_{\bar a}e^{\bar j}_{\bar b}e^{\bar k}_{\bar c}\,
\langle \psi{\,}^{d}\bar \psi{\,}^{\bar a} \bar \psi{\,}^{\bar b} \bar \psi{\,}^{\bar c}\rangle \,
(\partial_{m}\partial_{\bar i}\bar {\cal B}_{\bar j\bar k})\,.
\ee
 Here, $\langle \ldots \rangle$ denotes the Weyl-ordered products of fermions, $R$ is the standard scalar curvature
of the metric $h_{j\bar k}$, and $\triangle^{\rm cov}$ is the covariant Laplacian calculated with the
hatted affine and spin connections,
  \be
-\triangle^{\rm cov} \ =\ h^{\bar k j} \left( {\cal P}_j {\bar {\cal P}}_{\bar k} +
i \hat \Gamma^{\bar q}_{j \bar k} {\bar {\cal P}}_{\bar q}
+  {\bar {\cal P}}_{\bar k} {\cal P}_j + i \hat \Gamma^{s}_{\bar k j} {{\cal P}}_s \right),
 \ee
where ${\cal P}_M$ are still given by Eq.(\ref{calP}) with $P_M \to -i \partial_M$.

The structure $\sim R - CC$ entering the expression   (\ref{kvantH}) can be written in the real notation as
 ${\tilde R} = R - \frac 1{12} C_{MNP} C^{MNP}\,$, that is to be compared with the ``hatted'' counterpart of $R$ calculated
with non-symmetric affine connections $\hat \Gamma$,
 $$
\hat R = R - \frac 14\, C_{MNP}C^{MNP}\,.
$$

Let us say a few words about geometric interpretation of our system.

 By the same token as in the K\"ahler torsionless case \cite{Wipf,IS} (the extra term
 $\sim C \psi^3$ in the classical supercharge (\ref{Qreal}) does not create ordering problems),
the {\it sum} ${\cal Q}^{\rm cov}$ of the supercharges (\ref{Qcovgen}) can be interpreted as the Dirac
operator on the manifold equipped with the torsion $\frac 13 C_{MNP}$,
 \be
\lb{Qcovreal}
 {\cal Q}^{\rm cov} = Q^{\rm cov} + \bar Q^{\rm cov}\ \equiv \  i\gamma^M {\tilde \nabla}_M\,,
 \ee
where ${\tilde \nabla}_M  = \Pi_M - \frac i2 {\tilde \Omega}_{M, BC} \gamma^B \gamma^C$
and
  \be
{\tilde \Omega}_{M, BC} = \Omega_{M, BC}  + \frac 16 e_B^L e_C^K C_{LKM} \ .
 \ee

The {\it difference} $Q^{\rm cov} - \bar Q^{\rm cov}$ is isomorphic to the operator $ \gamma^M  I_M^N {\tilde \nabla}_N $, where
${\tilde \nabla}_M = \partial_M + \frac 12 {\tilde \Omega}_{M, AB} \gamma^A \gamma^B$ and
$I_N^M$ is the covariantly constant complex structure matrix ($I^2 = -1\,,$ $\nabla_T I_N^M =0$).

It was noticed in \cite{IS} that,
 with the vanishing extra torsion terms $\propto {\cal B}, \bar {\cal  B}$ and for a particular choice
$W =  \frac 14 \ln \det h$, the supercharges (\ref{Qcovgen}) realize the Dolbeault  complex. It involves the operator
of the exterior holomorphic derivative $\partial$ and its conjugate $\partial^\dagger$.

When $W =  -\frac 14 \ln \det h$, the supercharges are isomorphic to the operators $\bar \partial$ and $\bar \partial^\dagger$
of the anti-Dolbeault complex.
For other choices of $W$, they realize a twisted Dolbeault complex with  $\partial_{\cal A} = \partial - {\cal A}$,
where ${\cal A} = \partial W$ is an exact $(1,0)$- form that can be interpreted as a gauge field. The latter might be nontrivial.
The fact that $\partial {\cal A} = 0$ {\it does} not
mean that the {\it real} gauge field   $A_M = (-i\partial_j W, i\partial_{\bar j} W )$ has a zero curl.

 In a more general case we have considered in this section,
we are dealing with the torsion-deformed twisted Dolbeault complex \cite{twistDolb}
involving the operators [cf. Eq.(\ref{deformdd+}) ]
 \be
\lb{deformAB}
 \partial_{W, B} {\cal O}  &=& \partial {\cal O} - \partial W \wedge {\cal O}  - \partial {\cal B} \wedge {\cal O}\,, \nn[6pt]
 \partial_{W, B}^\dagger {\cal O}  &=& \partial^\dagger {\cal O} +
\langle \bar \partial W, {\cal O} \rangle -
 \langle \bar \partial \bar {\cal B} , {\cal O} \rangle \ .
 \ee
The notation $\langle X, Y \rangle$ stands now for the complex interior product. For example, if $X$ is a $(0,1)$-form and
$Y$ is a $(1,0)$-form, $\langle X, Y \rangle = h^{\bar j k } X_{\bar j} Y_k$.

Note that the quantum supercharges (\ref{Qcovgen}) and the Hamiltonian (\ref{Hcovgen}) depend on ${\cal B}$ only via
its exterior derivative $\partial {\cal B}$. This means that ${\cal B}$ is defined only up to a
{\it gauge transformation} ${\cal B} \to {\cal B} + \partial {\cal A}$.

\setcounter{equation}{0}
\section{Vacuum states}

In this section we will find the zero-energy vacuum states wave functions $\Phi_{\cal B}$
in the $({\bf 1, 2, 1})$ and $({\bf 2, 2, 0})$ SQM sigma-models with nonzero torsions
produced by the terms (\ref{realB2})-(\ref{Bterms}) with tensors ${\cal B}$. These wave functions
are solutions of the generic equations
\be
\label{QQbar}
Q^{\rm cov}\Phi_{\cal B}=0\,,\qquad \bar Q^{\rm cov}\Phi_{\cal B}=0\,.
\ee
As a result, the vacuum wave functions $\Phi_{\cal B}$ in the presence of torsions
will prove to be  certain deformations of the torsionless wave functions $\Phi_{{\cal B}{=}0}$,
thus encompassing the same number of states as $\Phi_{{\cal B}{=}0}$.
An important tool for obtaining the general solution for the vacuum wave functions $\Phi_{\cal B}$
will be the representation \p{Qcov-121} for the quantum supercharges,
where the terms with torsions (as well as the potential terms) are absorbed into a similarity transformation
of the ``undeformed'' supercharges. We will essentially exploit the geometric correspondence with the
de Rham complex in the $({\bf 1, 2, 1})$ case [such that  \p{Qcov-121} acquires the form
\p{ddefexp}] and the Dolbeault complexes in the $({\bf 2, 2, 0})$ case.
We will limit our analysis to the spheres $S^n$ and $\mathbb{CP}^n$ manifolds, in the first and the
second cases,
respectively.

\subsection{de Rham complex with torsions}

For the de Rham complex (i.e. for the $({\bf 1,2,1})$ SQM model of Sect. 2), the Witten index Tr$\{(-1)^F\}$
coincides with the
Euler characteristic $\chi$ of the manifold. Consider $S^n$ as the
simplest example.

 When $n$ is even, $\chi=2$, which suggests the presence of
two bosonic zero modes. When the torsions are absent, these zero modes are seen
explicitly - it is the constant 0-form and the volume $n$-form.\footnote{Such zero modes exist for any target
Riemann geometry, not only for spheres. In the generic case, other zero modes can be present.}
Witten index cannot change under a smooth deformation. This assures the presence
of two bosonic zero modes in the spectrum also for a deformed complex.

When $n$ is odd, the Euler characteristic vanishes. If the manifold has an isometry, one can consider
another index, the so called Lefshetz number Tr$\{(-1)^F K\}$, where $K$ is an isometry commuting
with the Hamiltonian, for example - a reflection of one of the coordinates.
 For a ``round''  odd-dimensional sphere,
this Lefshetz number is equal to 2, which means, again, the presence of two zero modes in the deformed complex
if the deformation respects this isometry \cite{Witgeom}.

For a ``crumbled'' sphere without any isometry
(or when the isometry is not respected by the deformation), this argument does not work. Still, one can
prove that the number of supersymmetric vacua is left unchanged.

The situation is especially simple for the deformation \p{deformW} where the deformed vacua can be
found explicitly. Indeed,
 the operator  $d_W = e^{W} d e^{-W} $  annihilates the 0-form $e^{W}$ (being 0-form, it is automatically annihilated
 by $d^\dagger_W = e^{-W} d^\dagger e^{W}\,$). Likewise, the operator
  $d_W^\dagger = e^{-W} d^\dagger e^{W} $  annihilates the form
  $e^{-W} {\cal V}_n = e^{-W} \sqrt{g} \, dx^1 \wedge \cdots \wedge dx^n\,$, which is also automatically annihilated by $d_W\,$.

To prove the same for the deformation \p{deformdd+}, a little more elaborate reasoning is required.
Let us first prove the {\it theorem of existence} - show that the equations \p{QQbar}
or, in the differential form language,
\be
\lb{eqvac}
d_{\cal B} \Phi_{\cal B} = d_{\cal B}^\dagger \Phi_{\cal B} = 0\,,
 \ee
have two nontrivial solutions. In fact, the proof goes in the same way, irrespective of whether $n$ is odd or even.

Consider the form $\Phi = e^{\cal B}$. From \p{ddefexp}, we immediately deduce that it is closed in the deformed sense,
$d_{\cal B} \Phi = 0$. However, it cannot be exact,
$\Phi \neq d_{\cal B} \Phi$. Indeed, the identity $e^{\cal B} = d_{\cal B} \Psi =  e^{\cal B} d e^{-\cal B}
\Psi$ would mean that $1 = d\left( e^{-{\cal B}} \Psi \right)$. But a 0-form cannot be $d$ - exact as
the operator $d$ increases the order of the form  by one \footnote{This argument is specific for $S^n$ (only which we study here), but the isomorphism of
the cohomologies of the twisted de Rham complex and the untwisted one is a quite general fact
\cite{MWu}. Indeed, using the property \p{ddefexp}, it is easy to observe that a form $\alpha$
is $d_{W,{\cal B}}$ - closed if and only if the form $e^{-W-{\cal B}} \alpha$ is $d$-closed and the form
  $\alpha$
is $d_{W,{\cal B}}$ - exact if and only if the form $e^{-W-{\cal B}} \alpha$ is $d$-exact.}.

Note now that {\it any} form of even order and, in particular, $\Phi$ can be represented as
 \be
\label{HodgeB}
\Phi = d_{\cal B} {\cal X} + d^\dagger_{\cal B} {\cal Y} + \Phi_{\cal B}  \, ,
 \ee
where ${\cal X}$ and ${\cal Y}$ are some odd-order forms, while the form $\Phi_{\cal B}$ satisfies \p{eqvac} and is thus $d_{\cal B}$ - harmonic.
A mathematician will recognize in \p{HodgeB} a variant of the Hodge decomposition theorem \cite{Hodge}. Its physical meaning
is rather transparent.  It simply says that the Hilbert space
of any SQM system is spanned by {\it (i)} zero modes of the Hamiltonian (i.e. $\Phi_{\cal B}$), {\it (ii)} the states annihilated
by the supercharge $Q$ but not by the supercharge $\bar Q$ (i.e. $ d_{\cal B} {\cal X}$) and {\it (iii)} the states annihilated
by $\bar Q$ but not by $Q$ (i.e. $ d^\dagger_{\cal B} {\cal Y}$).

For our form $\Phi$, annihilated by the action of $Q \equiv d_{\cal B}$, the second term in the expansion \p{HodgeB}
must be absent. As the form is not exact, there should be some nontrivial nonzero
$\Phi_{\cal B} = e^{\cal B} - d_{\cal B} {\cal X}$ (belonging to the same cohomology class as $e^{\cal B}$). This is a first
solution of \p{eqvac}.

To find the second solution, consider the volume form ${\cal V}_n$. It is $d$-closed and also $d_{\cal B}$ - closed.
It is the zero mode of the untwisted complex and hence cannot be $d$-exact. It follows that neither it is $d_{\cal B}$-exact.
Indeed, ${\cal V}_n = e^{\cal B} d e^{-\cal B} {\cal X}$ would mean that  $e^{-\cal B}{\cal V}_n =  {\cal V}_n =
d \left(e^{-\cal B} {\cal X} \right)$.
Using the same reasoning as above,
we derive that the form
 $  {\cal V}_n  $ can be presented as
 \be
 {\cal V}_n  \ =\ d_{\cal B} {\cal Z} + \tilde \Phi_{\cal B} \ ,
 \ee
where $\tilde \Phi_{\cal B}$ is a nontrivial
$d_{\cal B}$ - harmonic form of the same order as $  {\cal V}_n  $.
In the physical language, this is the second bosonic zero mode for even-dimensional spheres and a fermionic zero mode
for the odd-dimensional ones. This fermionic vacuum state is needed to compensate
the bosonic zero mode and so to ensure the vanishing Witten index in the case of odd-dimensional spheres.

The theorem is proven.

It is interesting, however, to find the solutions of the equation \p{eqvac} explicitly. It is possible
to  do this perturbatively to any
order of perturbation theory in ${\cal B}$. Let us see how it works.

The simplest nontrivial case is $S^3$ (for 2-manifolds, the deformation \p{deformdd+} vanishes). Let us look for the
solution to the equations \p{eqvac} in the form
 \be
\lb{AnsS34}
\Phi = 1 + Y_2\,,
\ee
where $Y_2$ is a 2-form (this Ansatz asserts that the undeformed function is just $\Phi_0 = 1$).
For $S^3$, Eq.(\ref{eqvac}) implies
 \be
\lb{eqsS3}
 S^3: \ \ \ \ \ \ \ \ \ d Y_2 = d {\cal B} \ , \ \ \ \ \ \ \
 d^\dagger Y_2 = 0\,.
 \ee
  The solution of the first equation in \p{eqsS3} is
\be
\lb{Y2}
Y_2={\cal B}+ d{\cal A}_1\,,
\ee
with an arbitrary 1-form ${\cal A}_1$. The latter
can be written as ${\cal A}_1 = d\omega_0 + d^\dagger \omega_2$, where $\omega_0$ and $\omega_2$ are, respectively,
 some 0- and 2-forms (the 0-form term is just a gauge freedom).
Such representation
is guaranteed by the Hodge decomposition theorem with respect to the usual de Rham complex $d, d^\dagger$, bearing
in mind that there are no zero-mode 1-form ---  the Betty number $\beta_1$ for $S^3$ vanishes.
The term $d\omega_0\,$, being a gauge freedom, does not affect the solution \p{Y2} and we are safe to disregard it and
{\it choose the gauge} ${\cal A}_1 = d^\dagger \omega_2$ such that $d^\dagger {\cal A}_1 = 0$. Then the second
equation in \p{Y2} yields
$$
\triangle {\cal A}_1 = -d^\dagger{\cal B} \, ,$$
 where $\triangle\equiv d d^\dagger+ d^\dagger d$ is the covariant Laplacian.
The latter can be inverted (again, we are exploiting the fact that it does not have zero modes in the Hilbert space of
1-forms), which gives the solution
 \be
\lb{solS3}
{\cal A}_1 = - \triangle^{-1} d^\dagger {\cal B}
 \ee
for any ${\cal B}$.

For $S^4$, we may seek for the solution in the same form (\ref{AnsS34}) as for $S^3$.
We obtain the same equations (\ref{eqsS3}) and the same solution \p{Y2}, \p{solS3}. Note that in this case
we {\it could} also add  some 4-form $Y_4$ in the Ansatz \p{AnsS34} but the
equations \p{eqvac} would imply that $Y_4 = 0\,$.

For $S^5$ and for $S^6$, the Ansatz (\ref{AnsS34})
is not compatible with the equations \p{eqvac} and we  are {\it forced} to extend it by adding a 4-form $Y_4$
\be
\lb{AnsS56}
\Phi = 1 + Y_2 + Y_4 \, .
\ee
Putting \p{AnsS56} in \p{eqvac}, we derive the following
equations for the forms $Y_{2,4}$:
    \be
S^{5,6}\,:\qquad \ba{ll}
&d Y_2 = d {\cal B} \ , \\[7pt]
&d Y_4 = d {\cal B}\wedge  Y_2\ ,
\ea\qquad\qquad
\ba{ll}
& d^\dagger Y_2 - \langle d \bar {\cal B}, Y_4 \rangle = 0 \ , \\[7pt]
&d^\dagger Y_4 =  0 \ .
\ea \lb{eqsS56}
\ee
A generic solution of the equations with the operator $d$ (left column of \p{eqsS56}) is
\be\lb{Y4}
&&Y_2={\cal B}+ d{\cal A}_1\,, \nn
&&Y_4={\textstyle\frac12}\,{\cal B}\wedge{\cal B}+{\cal B}\wedge d{\cal A}_1+ d{\cal A}_3\,,
\ee
where  ${\cal A}_3$ is an arbitrary  3-form defined up to
gauge transformations ${\cal A}_3\to{\cal A}_3+ d\omega_2$.
By the same token as above, we use this gauge freedom to choose the gauge ${\cal A}_3 = d^\dagger \omega_4\,$, in which
case $d^\dagger {\cal A}_3 = 0$. We also assume as before that $d^\dagger {\cal A}_1 = 0$ (by choosing the proper gauge).

Then the equations in right column of \p{eqsS56} amount to the following system
\begin{eqnarray}
 \triangle {\cal A}_1
 &=& -d^\dagger {\cal B} + \langle d \bar {\cal B}, {\textstyle\frac12}\,{\cal B}\wedge{\cal B} \rangle +
\langle d \bar {\cal B}, {\cal B}\wedge d{\cal A}_1 \rangle +
\langle d \bar {\cal B}, d{\cal A}_3  \rangle
\,,
\label{eq-A1}\\[6pt]
\triangle {\cal A}_3&=& -\frac 12 d^\dagger ({\cal B}\wedge{\cal B}) - d^\dagger\left({\cal B}\wedge d{\cal A}_1\right)
\,,\label{eq-A3}
\end{eqnarray}
which allows one to define ${\cal A}_1$ and ${\cal A}_3$.

 The solution to this set of equations can be found as a perturbation series
with respect to the torsion field ${\cal B}$. It is nothing but a standard quantum mechanical perturbation series,
which is, however, essentially simplified
due to supersymmetry. Indeed, the vacuum energy remains zero, and so we have to solve not the second order
Schr\"odinger equation, but rather the first order equations $Q\Phi = \bar Q \Phi = 0$.

Substitute in \p{eq-A1}, \p{eq-A3}  the formal expansions of ${\cal A}_{1,3}$ in ${\cal B}$. It turns out that ${\cal A}_1$
is expanded over {\it odd} powers of ${\cal B}$, while ${\cal A}_3$ - over {\it even} powers,
\begin{eqnarray}
{\cal A}_1&=&{\cal A}_1^{(1)}+{\cal A}_1^{(3)}+{\cal A}_1^{(5)}+{\cal A}_1^{(7)}+\cdots\,,\label{A1}\\
{\cal A}_3&=&{\cal A}_3^{(2)}+{\cal A}_3^{(4)}+{\cal A}_3^{(6)}+{\cal A}_3^{(8)}+\cdots\,.\label{A3}
\end{eqnarray}
The explicit solution can be found by iterations.
For example, from \p{eq-A1}, we find ${\cal A}_1^{(1)}=-{\triangle}^{-1} \,d^\dagger {\cal B}$.
Then \p{eq-A3} gives ${\cal A}_3^{(2)}=
-{\triangle}^{-1}\left(d^\dagger {\cal B}\wedge{\cal B} - d^\dagger
({\cal B}\wedge {\triangle}^{-1}\,d^\dagger {\cal B})\right)$.
After that, we find ${\cal A}_1^{(3)}$ from  \p{eq-A1} and then ${\cal A}_3^{(4)}$ from  \p{eq-A3} and
so on, step by step. We of course need to assume that this perturbation series is convergent and the resulting full
vacuum wave function is a regular form on the whole manifold, like ${\cal B}$ itself.

For $S^{7,8}$, we are obliged to include a 6-form in the Ansatz, $\Phi = 1 + Y_2 + Y_4+ Y_6$.
We obtain the following equations
\be \lb{eqsS78}
S^{7,8}\,:\qquad \ba{ll}
&d Y_2 = d {\cal B} \ , \\[7pt]
& d Y_4 = d {\cal B}\wedge  Y_2\ ,\\[7pt]
&d Y_6 = d {\cal B}\wedge  Y_4\ ,
\ea\qquad\qquad
\ba{ll}
& d^\dagger Y_2 - \langle d \bar {\cal B}, Y_4 \rangle = 0 \ , \\[7pt]
& d^\dagger Y_4 - \langle d \bar {\cal B}, Y_6 \rangle = 0 \ , \\[7pt]
&d^\dagger Y_6 =  0 \ .
\ea
\ee

A general solution of the equations in the left column is
\be\lb{Y6}
&&Y_2={\cal B}+ d{\cal A}_1\,, \nn
&&Y_4={\textstyle\frac12}\,{\cal B}\wedge{\cal B}+{\cal B}\wedge d{\cal A}_1+ d{\cal A}_3\,, \nn
&&Y_6={\textstyle\frac{1}{6}}\,{\cal B}\wedge{\cal B}\wedge{\cal B}+
{\textstyle\frac{1}{2}}\,{\cal B}\wedge{\cal B}\wedge d{\cal A}_1+{\cal B}\wedge d{\cal A}_3
+ d{\cal A}_5\,
\ee
with arbitrary  ${\cal A}_{1,3,5}$.
Choosing the gauge $d^\dagger {\cal A}_1= d^\dagger {\cal A}_3= d^\dagger {\cal A}_5=0$,
 using the expansions  \p{A1}, \p{A3} and
\begin{equation}\label{A5}
{\cal A}_5 ={\cal A}_5^{(3)}+{\cal A}_5^{(5)}+{\cal A}_5^{(7)}+{\cal A}_5^{(9)}+\cdots\,,
\end{equation}
we find all the components in the decompositions \p{A1}, \p{A3}, \p{A5}  step by step
from the equations in the right column in \p{eqsS78}, similarly to the $S^{5,6}$ case.

The solutions can be represented in the following nice form,
\begin{equation}\label{WF-Bq-com-odd}
\Phi_{\cal B} =e^{\mathcal{B}}\left(1  + d \mathcal{A}_{1}+ d \mathcal{A}_{3}+\ldots
+d \mathcal{A}_{n-2}\right)
\end{equation}
for odd-dimensional spheres and
\begin{equation}\label{WF-Bq-com-even}
\Phi_{\cal B} =\left[e^{\mathcal{B}}-{\textstyle\frac{1}{(n/2)!}}\,
\mathcal{B}^{\,n/2}\right]\left(1  +d \mathcal{A}_{1}+ d \mathcal{A}_{3}+\ldots
+d \mathcal{A}_{n-3}\right)
\end{equation}
for even-dimensional ones.

Up to now we only constructed the solution obtained  by a perturbation of the constant 0-form due to nonzero ${\cal B}$ .
But the same analysis can be done for the volume form ${\cal V}_n$ by duality. For example, for $S^7$, we can
start from the Ansatz
\be
\lb{AnsS3vol}
\Phi = {\cal V}_7  - \langle \bar Y_2 , {\cal V}_7 \rangle  + \langle \bar Y_4 , {\cal V}_7 \rangle  -
\langle \bar Y_6 , {\cal V}_7 \rangle \, ,
\ee
with  arbitrary $\bar Y_{2,4,6}$. The latter satisfy exactly the same equations as before, with the same solutions.

The solution \p{WF-Bq-com-odd} can be recast as $\Phi_{\cal B} = e^{\mathcal{B}} + d_{\cal B}
\left[  e^{\mathcal{B}} ({\cal A}_1 + \ldots +  \mathcal{A}_{n-2})\right]$, i.e. it belongs
to the cohomology class
 \be
\label{cohom}
\Phi_{\cal B} \ =\  e^{\mathcal{B}} + d_{\cal B} {\cal X} \, .
 \ee
 The solution \p{WF-Bq-com-even} can be presented as
$e^{\cal B}  - {\textstyle\frac{1}{(n/2)!}}\,
\mathcal{B}^{\,n/2} + d_{\cal B} \left[e^{\cal B} ({\cal A}_1 + \ldots +
\mathcal{A}_{n-2})\right]$. It belongs to a mixture of the cohomology class \p{cohom} and the class
 \be
\label{cohom1}
\hat{\Phi}_{\cal B} \ =\  {\cal V}_{n}  + d_{\cal B} \hat{\cal X} \, .
 \ee

The results \p{WF-Bq-com-odd} and \p{WF-Bq-com-even} can be easily translated into the ``physical'' notation.
For example, \p{WF-Bq-com-odd} describes the wave functions of the form
 \begin{equation}\label{WF-Bq-odd}
\Phi_{\cal B} =e^{\psi^M\psi^N\mathcal{B}_{MN}}\left(1  +\psi^M\psi^N \partial_M\mathcal{A}_{N}
+\ldots +
\psi^{M_1}\ldots\psi^{M_n}\partial_{M_1}\mathcal{A}_{M_2\ldots M_n}\right).
\end{equation}
This expression for the ground state wave function  matches well with the representation
\p{Qcov-121} for the supercharges.

\subsection{Dolbeault complex with torsions}

Consider first the Lagrangian \p{start} without the extra torsion terms. The number of vacuum states is given
by the Atiyah-Singer theorem. The latter is widely known when the manifold is K\"ahler and
the index of the Dolbeault operator coincides with the standard Dirac
index.\footnote{In the non-K\"ahler case, there are certain
complications, but the Atiyah-Singer theorem still can be formulated and proven. The physical proof was given
in a recent paper \cite{Hirzebruch}.}
For example, in the $\mathbb{CP}^n$ case with the additional
condition
 \be
\lb{Wcan}
 W \ =\ \frac {q}{2(n+1)} \ln \det h \ =\ -\frac q2 \ln (1 + \bar z z)
 \ee
(this choice of $W$ defines what
is called the {\it canonical} or {\it determinant} bundle), the index is equal to
 \be
\lb{ICPn}
I_{\mathbb{\mathbb{CP}}^n} \ =\ \left( \begin{array}{c} q + (n-1)/2 \\ n \end{array} \right),
 \ee
where $q$ must be integer for odd $n$ and half-integer
for even $n\,$. For other values of $q$, one cannot consistently define the Hilbert space, where
the spectrum of the Hamiltonian is supersymmetric \cite{flux}. This means that, in contrast to the real case, introduction
of the term $\sim W$ in the action cannot be considered as a smooth deformation, and this is the reason why
the index \p{ICPn} depends on $q$.

When $|q| < \frac {n+1}2$, the zero-energy vacuum states are absent, indicating the
spontaneous breaking of supersymmetry in this case\footnote{This in turn is related to spontaneous breaking
of supersymmetry  in $d=3$ supersymmetric Yang-Mills-Chern-Simons theory \cite{Witten,paradox}.}.
 When $q =  \pm \frac {n+1}2$,
we are facing the Dolbeault (respectively, anti-Dolbeault) complex and
there is only one vacuum state in the sector $(p,q) = (0,0)$ (respectively, in the sector  $(p,q) = (n,n)$).
For larger values of $|q|$, we are dealing with the twisted Dolbeault complex (with an additional gauge field) and
 there are several such states. Let us first discuss the pure Dolbeault complex with  $q = (n+1)/2$.
 The supercharges $Q, \bar Q$
can be interpreted in this case as the operators of exterior holomorphic derivative
and its Hermitian conjugate.

The torsion term \p{extra} can be introduced as a smooth deformation, and  the index cannot change.
Thus, one can expect the system (\ref{Qcovgen}) to have exactly the same number of vacuum states (\ref{ICPn}) as
in the case of ${\cal B} = 0$. These states can be constructed along the same lines as for the de Rahm
complex.

In simplest nontrivial cases, $\mathbb{CP}^3$, $q=2\,$, and $\mathbb{CP}^4$, $q=\frac 52$, the deformed
vacuum wave functions can be found {\it exactly}. Indeed, we can adopt the Ansatz $\Phi = 1 + Y_{(2,0)}$, where $Y_{(2,0)}$ is a (2,0)-form.
By analogy with \p{Y2}, \p{solS3}, the solution to the equations
$\partial_{\cal B} \Phi = \partial^\dagger_{\cal B} \Phi = 0$ is
$Y_{(2,0)} = {\cal B}  - \partial \triangle^{-1} \partial^\dagger {\cal B}$,
where $\triangle$
 is the Laplacian (for K\"ahler manifolds, there is only one covariant Laplacian, $\triangle =
\partial \partial^\dagger +   \partial^\dagger \partial    = \bar\partial \bar\partial^\dagger +
   \bar\partial^\dagger \bar\partial\,$).

The analysis for $\mathbb{CP}^{n}$ with $n>4$ repeats without changes the analysis given above for $S^n, n>4$ .
One should replace
$\langle d \bar {\cal B} , \cdot \rangle$  in all formulas by
$\langle \bar \partial \bar {\cal B}, \cdot \rangle $ and also substitute everywhere  $\partial, \partial^\dagger$
for $d, d^\dagger\,$. The solutions are
  \begin{equation}\label{CPnodd}
\Phi_{\cal B} =e^{\mathcal{B}}\left(1  + \partial \mathcal{A}_{(1,0)}+ \partial \mathcal{A}_{(3,0)}+\ldots
+\partial \mathcal{A}_{(n-2,0)}\right)
\end{equation}
for odd $n$ and
\begin{equation}\label{CPneven}
\Phi_{\cal B} =\left[e^{\mathcal{B}}-{\textstyle\frac{1}{(n/2)!}}\,
\mathcal{B}^{\,n/2}\right]\left(1  +\partial \mathcal{A}_{(1,0)}+ \partial \mathcal{A}_{(3,0)}+\ldots
+\partial \mathcal{A}_{(n-3,0)}\right)
\end{equation}
for even $n$. All $(n,0)$-form $\mathcal{A}_{(n,0)}$ can be defined as series in ${\cal B}$
after fixing the gauges $\partial_{W'}^\dagger {\cal A}_{(n,0)} = 0$.

Like for the de Rham complex, the presence of the multiplier $e^{\mathcal{B}}$
in the solution \p{CPnodd}  is rather  natural, bearing in mind the representation
$\partial_{\cal B} =  e^{\cal B} \partial e^{-\cal B}$. In the physical notations, it reads
(cf. \p{Qcov-121}):
\begin{equation}\label{Qcov-pr1}
Q^{\rm cov} =
e^{\psi^i\psi^k\mathcal{B}_{ik}} \,\, Q^{\rm cov}_{{\cal B}=0} \, \,e^{-\psi^i\psi^k\mathcal{B}_{ik}}\,,
\end{equation}
where $Q^{\rm cov}_{{\cal B}=0}$ is the supercharge \p{Qcovgen} without torsion terms. For convenience of the reader
who prefers the more traditional language to the language of differential forms (which is most convenient
and adequate, in our opinion), in Appendix we present solutions of the vacuum equations \p{QQbar} for some particular
$\mathbb{CP}^n$ cases with $q = \frac{n+1}{2}$, using an equivalent tensorial notation.

Note that {\it both} \p{CPnodd} and \p{CPneven} belong to the cohomology
class $e^{\mathcal{B}} + \partial_{\cal B} {\cal X}$.
For \p{CPnodd},
it is clear, and \p{CPneven} differs from \p{CPnodd} by the $(n,0)$ - form ${\cal B}^{n/2}$.
In contrast to an n-form for $S^n$,
it is exact, ${\cal B}^{n/2} = \partial {\cal Y}$. It is guaranteed by the Hodge decomposition theorem
for the untwisted complex, where the term
$\partial^\dagger {\cal Z}$ is absent because it is the highest holomorphic form and the zero modes are
absent because
$\beta_{(n,0)} = 0$ for $\mathbb{CP}^n$. We finally note that ${\cal Y}$ is also $\partial_{\cal B}$ - exact as,
for a $(n-1, 0)$ - form ${\cal Y}$,
$\partial {\cal Y} = \partial_{\cal B} {\cal Y}$.

When $q > \frac {n+1}2$, we are facing the twisted Dolbeault complex. The equations are then
somewhat more complicated.
 When ${\cal B} = 0$, the vacuum states $\Phi_0$ should be defined by the equation
 \be
\label{eqPhi0}
 \partial_{W'} \Phi_0 =
e^{W'} \partial \left( e^{-W'} \Phi_0 \right) = 0 \, ,
 \ee
where $W'$ is the superpotential renormalized by the shift
$q \,\to\, 2s:=q \,{-}\, \frac {n+1}2 $ (see \cite{IS} for details),
\be
\lb{Wprim}
 W' \ =\ -s \ln (1 + \bar z z ) \, .
 \ee
The solution is then
 \be
\label{Phi0}
 \Phi_0 \ =\ e^{W'} \, R(\bar z) \, ,
 \ee
where $R(\bar z)$ is a polynomial of $\bar z^j$ of the degree not higher than $2s$ (to keep normalizability of
\p{Phi0}) \cite{IMT}. The index \p{ICPn} is none other than a number of coefficients in this polynomial.

Consider the simplest nontrivial $\mathbb{CP}^3$ case. Seek for the deformed vacuum wave function in the form
$$
\Phi_{\cal B}
 =  (1 + Y_{(2,0)}) \Phi_0 \, .
$$
The equations for $Y_{(2,0)}$ are
 \be
\lb{eqY2W}
 \partial_{W'} (Y_{(2,0)} \Phi_0)   &=& \partial_{W'} ({\cal B} \Phi_0)\,, \nn
\partial_{W'}^\dagger \left(  Y_{(2,0)} \Phi_0 \right) &=& 0
 \ee
with $\partial_{W'}^\dagger = e^{W'} \partial^\dagger e^{-W'}$.
A generic solution of the first equation in \p{eqY2W} is
  \be
 \lb{solY2W}
Y_{(2,0)} \  =\ {\cal B} + \Phi_0^{-1} \partial_{W'} {\cal A}_{(1,0)} \, .
  \ee

By the Hodge decomposition theorem with respect to the operators  $\partial_{W'}, \ \partial_{W'}^\dagger$
(it is valid as the operators
$\partial_{W'}$ and  \ $\partial_{W'}^\dagger$ satisfy the standard ${\cal N}=2$ supersymmetry algebra)
and from the fact that no zero modes
of the Hamiltonian $H_{W'} = \partial_{W'}^\dagger \partial_{W'} + \partial_{W'} \partial_{W'}^\dagger$  exist
in the (1,0) sector, the
form   ${\cal A}_{(1,0)}$ can be represented as  ${\cal A}_{(1,0)} =
\partial_{W'} \omega^{(0,0)} + \partial_{W'}^\dagger \omega^{(2,0)}$.
Let us choose a gauge, in which the first term is absent, so that $\partial_{W'}^\dagger {\cal A}_{(1,0)} = 0$. Then the second equation
in \p{eqY2W} acquires the form
$$ H_{W'} {\cal A}_{(1,0)} \ =\ -  \partial_{W'}^\dagger  {\cal B} \, .$$
The Hamiltonian $H_{W'}$ is positive-definite in the sector of (1,0)-forms and so can be inverted. This gives us
the form ${\cal A}_{(1,0)}$ and the solution \p{solY2W}.

The solutions for $\mathbb{CP}^n$ manifolds with higher $n$ have the form \p{CPnodd}, \p{CPneven}, where
one should make the substitution $\partial {\cal O} \to \Phi_{0}^{-1}\partial_{W'} {\cal O}
= \partial (\Phi_{0}^{-1} {\cal O}) $.

\setcounter{equation}0
\section{Summary and outlook}

In this paper we have studied the models of torsionful ${\cal N}{=}\,2$ supersymmetric quantum mechanics based
on the supermultiplets $({\bf 1,2,1})$ and $({\bf 2,2,0})$. These models are more general than
those which were studied in the literature up to now. For instance, the general ${\cal N}{=}\,2$ model based
on a sum of the superfield Lagrangians (\ref{start}) and \p{extra} was considered before basically at the classical level, while
its quantum version, including the explicit form of the relevant ${\cal N}{=}\,2$ supercharges, was known
only for a few particular cases \cite{Mavra,IS}. Also, the quantum models associated with
the multiplet $({\bf 1,2,1})$ were known either for the basic action \p{SN=2} (plus the potential term \p{SW}),
or for its modification obtained by adding
the action \p{realB2} with the {\it real} torsion potential \cite{Kimura} and without including any higher-order terms
like \p{realB4}. It should be pointed out that torsions of a certain special form always appear for  non-K\"ahler complex sigma model
\cite{IS}, but in the present paper we were interested in the models involving some {\it extra} torsion terms in the Lagrangians and supercharges
that are not related to the bosonic target space metric.

In all considered cases we constructed the corresponding quantum ${\cal N}{=}\,2$ superalgebra.
The general prescription is the use of the Weyl-ordered supercharges with subsequent passing
to the covariant supercharges which act in the Hilbert space with the geometrically motivated inner product.
Knowing these quantum  supercharges and interpreting them in terms of de Rahm (in the $({\bf 1, 2, 1})$ case)
and Dolbeault (in the $({\bf 2, 2, 0})$ case) complexes allowed us to explicitly find the vacuum states in the considered
models and check that their number does not change after switching on the torsions.
Such invariance is ensured  by the index theorem.
 enforced by a simple mathematical argument that the cohomologies for the twisted de Rham complex and for the untwisted complex
are the same \cite{MWu}.
The explicit construction of these states (we did it in the framework of the perturbative expansion over the torsion
${\cal B}$) is a new result.

In this paper we have considered ${\cal N}{=}\,2, d=1$ supersymmetric models which are in one-to-one
correspondence with the Rham complex and the Dolbeault complexes (untwisted and twisted).
It is interesting to explore, along the same geometric lines, the sigma models associated with
various ${\cal N}{=}\,4$ supermultiplets.
First, the number of different off-shell ${\cal N}{=}\,4$ supermultiplets
is considerably larger than that of ${\cal N}{=}\,2$ supermultiplets, which could lead to more
opportunities for the geometrical treatment of the corresponding models in terms of various complexes.
In particular, we expect to recover the so called quaternionic Dolbeault complex (see, e.g., \cite{verb} and refs. therein)
within such a context. Second, these systems are much richer, and so they could require some new means for the construction of the
corresponding quantum theories.

Even in the ${\cal N}{=}\,2$ case, there exists a class of models which until now
were not well studied  and geometric interpretation of which is unknown. They are based on
the multiplets $({\bf 1, 2, 1})$ described by the superfield action in \p{SN=2} in which the metric
$g_{MN}(X^P)$ contains {\it both}
symmetric {\it and} antisymmetric parts. Though the existence of such ${\cal N}{=}\,2$ models was mentioned
in \cite{PaCo,GPS}, no special attention was paid to them afterwards.

Finally, in this paper we only studied sigma models with ${\cal N}{=}\,2$ supermultiplets of the same type.
Of interest are also the models in which different types of supermultiplets enter simultaneously.
In particular, it is worthwhile to consider the models with isometries, a part of which is gauged (see \cite{DI2}
for the $d=1$ gauging procedure).

We will try to address these issues in the future.

\section*{Acknowledgements}

\noindent
We are indebted to D. Speyer and M. Verbitsky for useful discussions.
S.F. \&\ E.I. acknowledge support from the RFBR
grants 09-01-93107, 11-02-90445, 12-02-00517 and
a grant of the IN2P3-JINR Programme for 2012. They would like to thank SUBATECH,
Universit\'{e} de Nantes, for the warm hospitality
in the course of this study. The work of E.I. was carried out
under the Convention N${}^{\rm o}$ 2010 11780.

\renewcommand\theequation{A.\arabic{equation}} \setcounter{equation}0
\section*{Appendix \quad  }
Sometimes it is perhaps useful to have a more explicit form of the equations for the vacuum wave functions \p{QQbar} and their solutions.
Here we present a few examples related to the ${\cal B}$-deformed $\mathbb{CP}^n$ models with the condition \p{Wcan} and $q = \frac{n+1}{2}$
(i.e. those associated with the untwisted Dolbeault complex).

In the deformed $\mathbb{CP}^3$ case with $q=2$
Eqs. \p{QQbar} for the wave function $\Phi = 1 + Y^{(2)}_{ik}\psi^i\psi^k$ amount to the following set of explicit equations
\be
&& \partial_{[i} (Y^{(2)} - {\cal B})_{kl]} = 0\,, \quad \Rightarrow \quad Y^{(2)}_{kl} =  {\cal B}_{kl} + \partial_{[k} A_{l]} \,, \label{a1} \\
&& h^{i\bar k}\,\partial_{\bar k} Y^{(2)}_{il} = 0\,. \label{a2}
\ee
Using the gauge freedom $A_l \rightarrow A_l + \partial_l\omega$, one can choose the gauge $h^{i\bar k}\partial_{\bar k}A_i = 0$ and reduce
\p{a2} to
\be
\Delta A_l = - 2h^{t\bar k}\partial_{\bar k}{\cal B}_{tl}\,, \lb{LB}
\ee
where $\Delta = h^{t\bar k}\nabla_t \partial_{\bar k}$ is the covariant Laplace-Beltrami operator. Then Eq. \p{LB} can be solved for $A_l$
in terms of $\partial_{\bar k}{\cal B}_{tl}\,$,
\be
A_l = -2\Delta^{-1}h^{t\bar k}\partial_{\bar k}{\cal B}_{tl}\,. \lb{Delta-1}
\ee

For the $\mathbb{CP}^4$ case, with $q = \frac{5}{2}$ and $\Phi = 1 + Y^{(2)}_{ik}\psi^i\psi^k +  Y^{(4)}_{iklm}\psi^i\psi^k\psi^l\psi^m$,
we have the following system
\be
&& \partial_{[i} (Y^{(2)} - {\cal B})_{kl]} = 0\,, \label{b1} \\
&& h^{i\bar k}\,\partial_{\bar k} Y^{(2)}_{il} + 12\,h^{i\bar k}\,h^{m\bar t}\,h^{n\bar j}\,\partial_{[\bar k}\,\bar{\cal B}_{\bar t \bar j]}\,
Y^{(4)}_{imnl} = 0\,, \label{b2} \\
&& h^{i\bar k}\,\partial_{\bar k} Y^{(4)}_{ilmn} = 0\,.\label{b3}
\ee
The last equation implies $\partial_{\bar k}Y^{(4)}_{ilmn} = 0$, whence
\be
Y^{(4)}_{ilmn} = 0 \label{b33}
\ee
and we are left with the same solution as in the $\mathbb{CP}^3$ case.

In the $\mathbb{CP}^5$ case, with $q = 3$ and $\Phi = 1 + Y^{(2)}_{ik}\psi^i\psi^k +  Y^{(4)}_{iklm}\psi^i\psi^k\psi^l\psi^m\,$,
Eqs. \p{b1} - \p{b3} are supplemented by the following new equation
\be
\partial_{[i}\,Y^{(4)}_{kjlm]} - \partial_{[\bar i}  {\cal B}_{kj}\,Y^{(2)}_{lm]} = 0\,,\lb{b4}
\ee
which can be solved as
\be
Y^{(4)}_{iklm} = \frac{1}{2}\,{\cal B}_{[ik}\,{\cal B}_{lm]} + {\cal B}_{[ik}\,\partial_l A_{m]} + \partial_{[i}\,\Omega_{klm]}\,, \lb{b5}
\ee
where $\Omega_{klm}$ is a new totally antisymmetric function with its own gauge freedom $\Omega_{klm}
\rightarrow \Omega_{klm} + \partial_{[k}\omega_{lm]}$. Eqs. \p{b2} and \p{b3} can be used to solve for $A_m$ and $\Omega_{klm}$
in terms of ${\cal B}_{ik}, \bar{\cal B}_{\bar i \bar k}\,$ as perturbation series with respect to these fields. For instance,
in the lowest order $A_l$ is still given by the expression \p{Delta-1}, while $\Omega_{klm}$ subjected to the gauge condition
$h^{i\bar k}\partial_{\bar k}\Omega_{ilm} = 0$ is determined from Eq. \p{b3} as
\be
\Omega_{lmn} = -2\Delta^{-1}h^{i\bar k}\partial_{\bar k}\Big[{\cal B}_{[il}{\cal B}_{mn]} -
2{\cal B}_{[il}\partial_m\Delta^{-1} h^{t\bar p}\partial_{\bar p}{\cal B}_{tn]}\Big].
\ee

The solutions \p{a1}, \p{b33} and \p{b5} nicely match with the general formulas \p{CPnodd} and \p{CPneven}.

\end{document}